%% file: main.tex
\documentclass[a4paper,11pt]{article}

\input{macros}

\usepackage{geometry}
\usepackage{caption}
\usepackage{verbatim}
\usepackage{amsmath}
\usepackage{amsfonts}   
\usepackage{amssymb}    
\usepackage[utf8]{inputenc} 
\usepackage{lastpage}
\usepackage{fancyhdr}
\usepackage{graphicx}  
\usepackage{color}     
\usepackage{enumitem}
\usepackage{cite}      
\usepackage{hyperref}  
\usepackage{longtable}
\usepackage{listings}
\usepackage{bold-extra}
\usepackage{mathrsfs}
\usepackage[percent]{overpic}
\usepackage{amsthm}

\newcommand{\ignore}[1]{}

\date{\today}
\author{
    Yossi Azar\thanks{Blavatnik School of Computer Science,
    Tel Aviv University, Israel. Emails:
    \texttt{azar@tau.ac.il},
    \texttt{itamarbiran@mail.tau.ac.il}.}
    \qquad
    Itamar Biran\footnotemark[1]
}

\title{Competition Complexity of Min-Prophet Inequalities}

\begin{document}

\maketitle
\input{sections/abstract}
\newpage
\input{sections/introduction}

\input{sections/related_work}
\input{sections/organization}
\input{sections/preliminaries}
\input{resource_augmentation/lower_bound_adversarial_order}

\input{resource_augmentation/Algorithm_for_random_order}

\input{resource_augmentation/lower_bound_for_random_order}

\input{resource_augmentation/algorithm_for_loglogM_interleaved_variables}
\input{resource_augmentation/Algortihm_i.i.d_loglog}

\input{resource_augmentation/Block_model_Lower_Bound}

\input{sections/Conclusion}
\bibliography{references}
\appendix
\input{Appendix/Proof_Adverserial_lower_bound}
\end{document}

%% file: macros.tex
\usepackage{typearea}
\usepackage{setspace}

\usepackage{fullpage}

\usepackage{comment} 
\usepackage{bbm}
\usepackage{framed} 
\usepackage{url} 
\usepackage{complexity}
\usepackage{booktabs}
\usepackage{amsmath,amssymb}
\usepackage{float}

\usepackage{amsthm}
\usepackage{thmtools} 
\usepackage{thm-restate}
\usepackage{nicefrac}
\usepackage{calc}
\usepackage{enumerate}
\usepackage{enumitem}
\usepackage[usenames,dvipsnames]{xcolor}

\usepackage[ruled,vlined,linesnumbered]{algorithm2e}
\usepackage{forloop}

\usepackage{graphicx}
\usepackage[font=footnotesize,labelfont=bf]{subcaption}
\usepackage[font=footnotesize,labelfont=bf]{caption}
\usepackage[nobreak=true]{mdframed}
\usepackage{appendix}
\usepackage[noend]{algpseudocode}
\usepackage[colorinlistoftodos]{todonotes}
\usepackage{xr}
\usepackage{array}
\usepackage{xspace}
\definecolor{ForestGreen}{rgb}{0.1333,0.5451,0.1333}
\definecolor{DarkRed}{rgb}{0.8,0,0}
\definecolor{Red}{rgb}{1,0,0}
\usepackage[linktocpage=true,
pagebackref=true,colorlinks,
linkcolor=DarkRed,citecolor=ForestGreen,
bookmarks,bookmarksopen,bookmarksnumbered]{hyperref}

\usepackage[capitalise]{cleveref}
\crefrangelabelformat{enumi}{#3#1#4--#5#2#6}
\usepackage{parskip}
\usepackage{chngcntr}
\usepackage{mathtools,stackengine}
\usepackage{multirow}

\stackMath
\newcommand{\stackGeq}[1]{%
	\setbox0=\hbox{${}\mathrel{\stackon[-1pt]{\geq}{\scriptstyle\text{#1\strut}}}{}$}
	\xdef\tmpwd{\dimexpr\the\wd0\relax}
	\kern.5\tmpwd\mathclap{\box0}&\kern.5\tmpwd
}

\usepackage{nameref}

\usepackage{tikz}
\usetikzlibrary{patterns}

\allowdisplaybreaks

\DeclareMathOperator*{\expectation}{\mathbb{E}}

\newcommand{\expect}{\expectation\expectarg}
\DeclarePairedDelimiterX{\expectarg}[1]{[}{]}{%
	\ifnum\currentgrouptype=16 \else\begingroup\fi
	\activatebar#1
	\ifnum\currentgrouptype=16 \else\endgroup\fi
}

\DeclarePairedDelimiterX{\nicesetarg}[1]{\{}{\}}{%
	\ifnum\currentgrouptype=16 \else\begingroup\fi
	\activatebar#1
	\ifnum\currentgrouptype=16 \else\endgroup\fi
}

\newcommand{\innermid}{\nonscript\;\delimsize\vert\nonscript\;}
\newcommand{\activatebar}{%
	\begingroup\lccode`\~=`\|
	\lowercase{\endgroup\let~}\innermid 
	\mathcode`|=\string"8000
}

\newcommand\opt{\textsc{Opt}\xspace}

\newcommand\alg{\textsc{Alg}\xspace}

\counterwithin{equation}{section}

\usepackage{eqparbox}

\theoremstyle{plain}

\newtheorem{theorem}{Theorem}[section]

\newtheorem{lemma}[theorem]{Lemma}

\newtheorem{claim}[theorem]{Claim}

\newlength{\continueindent}
\usepackage{etoolbox}
\makeatletter
\newcommand*{\ALG@customparshape}{\parshape 2 \leftmargin \linewidth \dimexpr\ALG@tlm+\continueindent\relax \dimexpr\linewidth+\leftmargin-\ALG@tlm-\continueindent\relax}
\apptocmd{\ALG@beginblock}{\ALG@customparshape}{}{\errmessage{failed to patch}}
\makeatother

\makeatletter
\def\thm@space@setup{%
	\thm@preskip=\parskip \thm@postskip=0pt
}
\makeatother

\usepackage{etoolbox}
\usepackage{tikz}
\usetikzlibrary{tikzmark}
\usetikzlibrary{calc}

\errorcontextlines\maxdimen

\newcommand{\ALGtikzmarkcolor}{black}
\newcommand{\ALGtikzmarkextraindent}{4pt}
\newcommand{\ALGtikzmarkverticaloffsetstart}{-.5ex}
\newcommand{\ALGtikzmarkverticaloffsetend}{-.5ex}
\makeatletter
\newcounter{ALG@tikzmark@tempcnta}

\newcommand\ALG@tikzmark@start{%
	\global\let\ALG@tikzmark@last\ALG@tikzmark@starttext%
	\expandafter\edef\csname ALG@tikzmark@\theALG@nested\endcsname{\theALG@tikzmark@tempcnta}%
	\tikzmark{ALG@tikzmark@start@\csname ALG@tikzmark@\theALG@nested\endcsname}%
	\addtocounter{ALG@tikzmark@tempcnta}{1}%
}

\def\ALG@tikzmark@starttext{start}
\newcommand\ALG@tikzmark@end{%
	\ifx\ALG@tikzmark@last\ALG@tikzmark@starttext
	\else
	\tikzmark{ALG@tikzmark@end@\csname ALG@tikzmark@\theALG@nested\endcsname}%
	\tikz[overlay,remember picture] \draw[\ALGtikzmarkcolor] let \p{S}=($(pic cs:ALG@tikzmark@start@\csname ALG@tikzmark@\theALG@nested\endcsname)+(\ALGtikzmarkextraindent,\ALGtikzmarkverticaloffsetstart)$), \p{E}=($(pic cs:ALG@tikzmark@end@\csname ALG@tikzmark@\theALG@nested\endcsname)+(\ALGtikzmarkextraindent,\ALGtikzmarkverticaloffsetend)$) in (\x{S},\y{S})--(\x{S},\y{E});%
	\fi
	\gdef\ALG@tikzmark@last{end}%
}

\apptocmd{\ALG@beginblock}{\ALG@tikzmark@start}{}{\errmessage{failed to patch}}
\pretocmd{\ALG@endblock}{\ALG@tikzmark@end}{}{\errmessage{failed to patch}}
\makeatother

\algblock[with]{With}{EndWith}
\algblockdefx[With]{With}{EndWith}%
[1]{\textbf{with} #1 \textbf{do}}%
{}

\makeatletter
\ifthenelse{\equal{\ALG@noend}{t}}%
{\algtext*{EndWith}}
{}%
\makeatother

%% file: sections/abstract.tex
\begin{abstract}
We initiate the study of competition complexity for the single-choice
min-prophet problem, a minimization analogue of the resource-augmented
prophet inequalities recently studied for maximization. In this setting,
the online algorithm is given $k$ independent copies of an instance and must
select one variable, while the benchmark is the offline minimum on a single
copy. The $(1+\epsilon)$-competition complexity is the smallest $k$ for which
the online algorithm achieves expected cost at most $(1+\epsilon)$ times the
expected offline optimum. Unlike the maximization setting, the min-prophet
problem admits no constant competitive ratio even for i.i.d.\ instances,
making resource augmentation essential.

We study this question under three standard arrival models and uncover a
sharp separation in the amount of augmentation required. In the block model,
we show that the $(1+\epsilon)$-competition complexity is
$\Theta\left(\log\log\frac{M}{\epsilon}\right)$, whereas in the prophet
secretary model, where copies arrive in uniformly random order, it is
$\Theta\left(\log\frac{M}{\epsilon}\right)$. Here $M$ is the ratio between
the minimum expected cost of a single variable and the expected global
minimum. In contrast, under fully adversarial arrivals, no finite number of
copies suffices to guarantee any constant competitive ratio. Thus, while
resource augmentation can overcome the fundamental impossibility of
min-prophet under structured arrival models, it cannot do so under
adversarial ordering.
\end{abstract}

%% file: sections/introduction.tex
\section{Introduction}

The classical prophet inequality concerns the following online selection
problem. Let $X_1,\ldots,X_n$ be independent nonnegative random variables
drawn from known distributions $F_1,\ldots,F_n$, and revealed sequentially
to an online algorithm. Upon observing $X_i$, the algorithm must decide
immediately and irrevocably whether to accept it and terminate, or to reject
it and continue. Its performance is compared with that of a prophet who
knows all realizations in advance and therefore obtains
$\expect{\max_{i\in[n]} X_i}$. Classical results show that, for arbitrary
independent distributions, an online algorithm can obtain at least one half
of the prophet's expected reward
\cite{KrengelSucheston1977,KrengelSucheston1978}. In the i.i.d.\ setting, the
optimal worst-case guarantee is substantially higher and converges to
approximately $0.745$ \cite{KERTZ198688}.

A recent line of work has studied the \emph{competition complexity} of the
single-choice max-prophet inequality
\cite{BrustleCDEFV26,BrustleEtAl2024,ezrasec2024,Ezracor25}. Rather than
accepting the loss caused by online uncertainty, the algorithm is given
access to multiple independent copies of the instance. Competition
complexity asks how many copies are sufficient for an online algorithm to
obtain a $(1-\epsilon)$-approximation to the expected value obtained by a
prophet who observes one copy and selects its maximum. This question has
been studied under several arrival models, including the i.i.d.\ setting
\cite{BrustleEtAl2024}, the block model
\cite{BrustleCDEFV26}, the prophet secretary model
\cite{ezrasec2024}, and fully adversarial arrivals
\cite{BrustleCDEFV26}.

A natural question is whether the same resource-augmentation framework can
be applied to minimization. At first sight, this appears to be a symmetric
problem: replace the maximum by a minimum and ask for the number of copies
needed to approximate the offline optimum. However, the minimization
setting has a fundamental asymmetry. In the \emph{min-prophet} (or
\emph{cost-prophet}) inequality \cite{LivanosMehta2022}, the variables
represent costs, and the online algorithm must eventually select one of
them. Rejecting every variable is not an option: failing to stop incurs
cost $\infty$. Thus, unlike in maximization, the algorithm must not only
identify a favorable realization, but also guarantee that it makes a
selection at all.

This difference leads to qualitatively different online behavior. While
the max-prophet inequality admits a constant competitive ratio for arbitrary
independent distributions, Esfandiari et al.~\cite{EsfandiariEtAl2015}
showed that the competitive ratio of the min-prophet inequality is
unbounded even for i.i.d.\ instances. Consequently, the usual competitive
analysis cannot provide a meaningful constant-factor guarantee in this
setting. This makes resource augmentation particularly natural: can
additional independent copies overcome this impossibility, and how many
copies are required to bring the online cost arbitrarily close to the
prophet's cost?

Consider, for example, a traveler who needs to purchase a flight ticket before a given deadline and observes a sequence of prices over time. Purchasing a ticket too early may result in missing a cheaper offer, whereas waiting too long may force the traveler to purchase a ticket at a substantially higher price. A natural way to mitigate this risk is to start searching earlier, thereby increasing the number of purchasing opportunities. From this perspective, competition complexity quantifies how much additional flexibility is needed to compensate for the lack of knowledge of future prices.
In this work, we initiate the study of \emph{competition complexity for the
single-choice min-prophet inequality}. We study the $(1+\epsilon)$-competition complexity: the minimum number of copies needed for an online
algorithm to achieve expected cost at most $(1+\epsilon)$ times the
expected cost of a prophet that observes one copy and selects its minimum.
We consider three canonical arrival models: the block model, the prophet
secretary model, and the fully adversarial model. For arbitrary independent
distributions, we establish tight bounds in the block and prophet secretary
models, and a strong impossibility result under fully adversarial arrivals.

Our bounds are stated in terms of the accuracy parameter $\epsilon$ and the
instance parameter $M$, introduced by Azar et al.~\cite{azar2026}. Formally,
\[
M
=
\frac{\min_{j\in[n]}\expect{X_j}}
{\expect{\min_{j\in[n]}X_j}}.
\]
Thus, $M$ measures the gap between the expected cost of the best
fixed variable and the expected global minimum. From an economic
perspective, $M$ captures the potential benefit of comparing
realized prices rather than committing in advance to a particular
supplier. When $M$ is close to $1$, committing to the supplier
with the lowest expected cost is already nearly optimal. In
contrast, a large $M$ indicates that substantial savings can be
obtained by exploiting the variability in realized costs. Our
results quantify how many additional purchasing opportunities
are needed to achieve near-optimal costs without knowing
future prices.
For i.i.d.\ instances, this simplifies to
\[
M
=
\frac{\expect{X_i}}
{\expect{\min_{i\in[n]} X_i}},
\]
so $M$ measures how much better, in expectation, the minimum of $n$ draws
is than a single draw.

\subsection{Our results}
We study the single-choice min-prophet inequality through the lens of
competition complexity. For $\epsilon\in(0,1)$, we characterize the minimum
number $k$ of independent copies of the original instance that are necessary
and sufficient for an online algorithm to achieve a $(1+\epsilon)$-competitive
ratio. Our bounds depend on the parameter $M$.
Table~\ref{tab:max-min-comparison} summarizes our results and compares them
with the corresponding competition-complexity bounds for the max-prophet
problem.

\begin{table}[H]
    \centering
    \renewcommand{\arraystretch}{1.25}
    \begin{tabular}{@{}lcc@{}}
        \toprule
        \textbf{Setting}
        & \textbf{Max-prophet}
        & \textbf{Min-prophet} \\
        \midrule
        i.i.d.
        & \(\Theta\left(\log\log\frac1\epsilon\right)\)
          \cite{BrustleEtAl2024}
        & \(\Theta\left(\log\log\frac{M}{\epsilon}\right)\) \\

        Block arrival
        & \(\Theta\left(\log\log\frac1\epsilon\right)\)
          \cite{BrustleCDEFV26}
        & \(\Theta\left(\log\log\frac{M}{\epsilon}\right)\) \\

        Random arrival
        & \(\Theta\left(\sqrt{\log\frac1\epsilon}\right)\)
          \cite{ezrasec2024}
        & \(\Theta\left(\log\frac{M}{\epsilon}\right)\) \\

        Fully adversarial arrival
        & \(\Theta\left(\frac1\epsilon\right)\)
          \cite{BrustleCDEFV26}
        & Unbounded \\
        \bottomrule
    \end{tabular}
    \caption{Comparison between the \((1-\epsilon)\)-competition complexity
    of the max-prophet problem and the \((1+\epsilon)\)-competition complexity
    of the min-prophet problem.}
    \label{tab:max-min-comparison}
\end{table}

We obtain tight bounds in the three standard arrival models:
\begin{itemize}
    \item \textbf{Block model.}
    The $(1+\epsilon)$-competition complexity is
    \[
        \Theta\left(\log\log\frac{M}{\epsilon}\right).
    \]
    The upper bound holds for arbitrary independent distributions, whereas the matching lower bound already holds for i.i.d.\ instances. We also provide a
constant-competitive algorithm that uses \(O(\log\log M)\) blocks and requires
only the expectations of the individual variables and of their minimum.

 \item \textbf{Prophet secretary model.}
 The $(1+\epsilon)$-competition complexity is 
 \[
\Theta\left(\log\frac{M}{\epsilon}\right).
 \]
The upper bound holds for arbitrary distributions while the lower bound holds for large enough $M$.
\item \textbf{Adversarial-order model.}
    No finite number of copies guarantees any constant competitive ratio.
    More precisely, for every number of copies $k$ and every constant
    $C>0$, there exists an instance and an adversarial arrival order for which
    every online algorithm has a competitive ratio greater than $C$ independent of the number of copies used (for non constant $M$).
\end{itemize}

Note that for i.i.d.\ instances, the block-model upper bound together with the i.i.d.\ lower bound of Section \ref{sec:block-lower-bound} yields the same tight bound
\(\Theta(\log\log(M/\epsilon))\).

\subsection{Techniques}
\paragraph{Block model.}
For the upper bound, we use a dual algorithm of Brustle et al.~\cite{BrustleCDEFV26} for the max-prophet problem. The proof is done by a reduction to that block-threshold
algorithm, which achieves a
\((1-\delta)\)-approximation for the max-prophet problem using
\(O(\log\log(1/\delta))\) blocks, together with an extra special phase in our algorithm. To make this reduction work, we define appropriate variables for the
maximization setting. We show that with high probability the two algorithms make essentially the same selection. 
Choosing the parameters appropriately then
gives an \(O(\log\log(M/\epsilon))\)-block algorithm.

For the lower bound, we construct a family of heavy-tailed i.i.d.\
distributions and analyze the optimal online stopping rule on these
instances. We show that even obtaining a constant competitive ratio requires
\(\Omega(\log\log M)\) blocks. We then transform this construction to obtain
a \(\Omega(\log\log(1/\epsilon))\) lower bound for achieving a
\((1+\epsilon)\)-competitive ratio. Together, these arguments yield the
matching \(\Omega(\log\log(M/\epsilon))\) bound.

\paragraph{Prophet secretary model.}
For the upper bound, we use a single-threshold algorithm until the last copy
of the variable with minimum expectation arrives, which is accepted if the
algorithm has not stopped earlier. The probability of seeing no value below
the threshold decreases exponentially with the number of copies, while the
fallback value has expectation at most \(M\) times the prophet's value. This
gives an \(O(\log(M/\epsilon))\) upper bound. For the lower bound, we construct
an instance in which an unfavorable random order occurs with sufficiently
large probability and forces every algorithm to incur a cost higher than
\((1+\epsilon)\) times the prophet's cost.
\paragraph{Adversarial-order model.}
We use two types of variables: one is deterministically equal to \(M\), while
the other is \(0\) with high probability but has an arbitrarily large
expectation. By placing all deterministic variables first, the adversary
forces the algorithm either to accept \(M\) or to continue to the risky
variables, yielding an unbounded competitive ratio.

%% file: sections/related_work.tex
\subsection{Related work}

No previous work has addressed the competition complexity of the min-prophet. Hence, we first describe related work for the min-prophet and then discuss the competition complexity for the maximization problem.

\paragraph{Min-prophet inequalities.}
Early work by Assaf, Goldstein, and Samuel-Cahn
\cite{Assaf_Goldstein_Samuel-Cahn_2004} studied a two-choice
minimization stopping problem for i.i.d.\ variables. The difficulty of
the single-choice min-prophet inequality was later demonstrated by
Esfandiari et al.\ \cite{EsfandiariEtAl2015}, who showed that no
constant competitive ratio is possible even in the i.i.d.\ setting.

Livanos and Mehta \cite{LivanosMehta2022} gave a systematic study of the
i.i.d.\ min-prophet inequality. They identified conditions under which
the optimal stopping policy achieves a distribution-dependent constant
competitive ratio, proved a factor of at most \(2\) for
monotone-hazard-rate distributions, and exhibited distributions for
which the competitive ratio is unbounded. They also obtained
polylogarithmic guarantees for single-threshold algorithms. In subsequent
work, Livanos and Mehta \cite{Livanos2024} used extreme-value theory to
characterize the asymptotic competitive ratio for broad classes of
i.i.d.\ distributions and extended their analysis to single-threshold
and multi-unit variants.

More recently, Li and Wu \cite{LiWu2026} studied the min-prophet
inequality for distributions supported on $[1,b]$. They obtained
a competitive ratio of $b^{(1-1/n)^n}$ in the i.i.d.\ setting,
even when the underlying distribution is unknown, and a tight
competitive ratio of $\Theta(\sqrt{b})$ in the non-i.i.d.\
setting. Note that for distributions supported on $[1,b]$, we
have $M\leq b$. The parameter $M$ thus provides a more refined
measure of the instance and allows us to capture distributions
with unbounded support.

Related procurement models have also been considered. Qin, Vardi, and
Wierman \cite{Junjie24} study a fractional variant in which the decision
maker may procure portions from several arriving suppliers. Azar et
al.\ \cite{azar2026} introduced the parameter \(M\) used in the
present paper and studied a multi-choice min-prophet problem in which the
algorithm may make several selections and pays their minimum. In that
model, the augmented resource is the number of selections. The present
work instead retains the single-choice requirement and augments the
number of independent copies.

\paragraph{Competition complexity for max-prophet.}
The notion of competition complexity originates in the work of Bulow and
Klemperer \cite{BulowKlemperer1996}, who showed that, in a classical
auction setting, recruiting an additional bidder can be more valuable
than using a more sophisticated mechanism. More recently, this
resource-augmentation perspective has been applied to prophet
inequalities.

In the i.i.d.\ setting, the optimal asymptotic competitive ratio in the
classical, unaugmented max-prophet problem is approximately \(0.745\)
\cite{HillK82,Correa21}. Brustle et al.\
\cite{BrustleEtAl2024} studied its competition complexity and showed
that the exact competition complexity is unbounded, whereas the
\((1-\epsilon)\)-competition complexity is
\(\Theta(\log\log(1/\epsilon))\).

For independent, non-identically distributed variables in a fixed
order, the classical prophet inequality admits the tight \(1/2\)
guarantee of Krengel and Sucheston
\cite{KrengelSucheston1977,KrengelSucheston1978}. In the corresponding
block model, Brustle et al.\ \cite{BrustleCDEFV26} showed that
block-threshold algorithms are optimal and achieve
\(\Theta(\log\log(1/\epsilon))\) competition complexity, whereas
single-threshold algorithms require
\(\Theta(\log(1/\epsilon))\) blocks. They also studied the robustness of
these guarantees under bounded displacement and adversarial arrival
orders.

Under uniformly random arrival, the classical prophet secretary problem
was introduced by Esfandiari et al.\ \cite{EsfandiariEtAl2015} and has
since been studied extensively
\cite{Eshani2024,Azar18,Correa21,GioSec24,ChenSec25}.
Ezra and Garbuz \cite{ezrasec2024} studied its competition complexity
and showed that its \((1-\epsilon)\)-competition complexity is
\(\Theta(\sqrt{\log(1/\epsilon)})\). Thus, random-order arrival requires
asymptotically more copies than the block model.
Ezra and Garbuz \cite{Ezracor25} further extended the
competition-complexity framework to correlated rewards. They showed
that the required augmentation may depend on the number of variables
and that block-threshold algorithms need not have finite competition
complexity. More recently, Cruz-Ossa et al. \cite{cruzossa2026} studied competition complexity for multiple-selection i.i.d. prophet inequalities, characterizing the augmentation required by single-threshold algorithms when the algorithm and the prophet may select multiple values.

All these competition-complexity results for prophet inequalities
concern reward maximization. We study the corresponding question for
cost minimization. In the independent maximization setting, the optimal
bounds depend only on \(\epsilon\), whereas the competition complexity
of the min-prophet inequality necessarily also depends on the parameter
\(M\).

%% file: sections/organization.tex
 \subsection{Organization}
The remainder of the paper is organized as follows. In Section \ref{sec:preliminaries}, we formally define the single-choice min-prophet inequality, the resource-augmented arrival models, and the parameter $M$. In Section \ref{lower bound adversarial order}, we present our strong impossibility result under adversarial arrivals, demonstrating that no finite-resource augmentation can guarantee a constant competitive ratio; the full proof of this result is deferred to Appendix \ref{app: proof adversarial}. Next, we shift to the prophet secretary (random-order) model: Section \ref{Algortihm random order} introduces our $(1+\epsilon)$-competitive algorithm, followed by its corresponding lower bound in Section \ref{lower bound random order}. Finally, Sections \ref{constant block algorithm}-\ref{sec:block-lower-bound} concern the block model. Section \ref{constant block algorithm} presents a simple constant competitive algorithm, Section \ref{block algorithm} details our online algorithm achieving a $(1+\epsilon)$-competitive ratio, and Section \ref{sec:block-lower-bound} establishes the matching lower bounds with respect to both $M$ and $\epsilon$.

%% file: sections/preliminaries.tex
\section{Preliminaries}
\label{sec:preliminaries}

In this section, we formally define the single-choice min-prophet inequality, the parameter $M$, competition complexity, the resource-augmented input, and the arrival models considered in this work. Throughout the paper, all logarithms are base $2$.
\paragraph{The min-prophet inequality.}
An instance consists of independent nonnegative random variables
\(X_1,\ldots,X_n\), drawn from known distributions
\(F_1,\ldots,F_n\). We denote by \(ALG\) the value selected by the
online algorithm and by
\[
    OPT=\min_{i\in[n]}X_i
\]
the value selected by the prophet. The algorithm must select one of the
observed variables; otherwise, we set \(ALG=\infty\). An algorithm is
\(C\)-competitive if
\[
    \expect{ALG}\leq C\expect{OPT}.
\]
Unless stated otherwise, the variables are independent but not necessarily
identically distributed.  An instance is called \emph{i.i.d.} if
\[
    F_1=F_2=\cdots=F_n.
\]
The i.i.d.\ assumption is a restriction on the distributions and is the same for any of the arrival models defined below.

\paragraph{Continuity.}
For ease of exposition, we assume that the relevant distributions are
continuous. This is without loss of generality: in the presence of atoms,
the algorithm may use independent auxiliary randomness to break ties, or
equivalently randomize its decision at a threshold so as to realize any
desired quantile. We use this standard convention throughout.
\paragraph{The parameter \(M\).}
For an instance \(\mathbf{F}=(F_1,\ldots,F_n)\), we define
\[
    M(\mathbf{F})
    =
    \frac{\min_{j\in[n]}\expect{X_j}}
         {\expect{\min_{j\in[n]}X_j}}.
\]
When the instance is clear from context, we simply write \(M\).
We assume that
\[
    \min_{j\in[n]}\expect{X_j}>0.
\]
Otherwise, since the variables are nonnegative, some variable is equal to
\(0\) almost surely, and the problem is trivial. 
We only assume one variable has finite expectation; all other variables may have infinite expectation.

\paragraph{Competition complexity.}
For \(k\geq 1\), the online algorithm observes \(k\) independent copies of
the original instance. We denote the \(\ell\)-th copy of \(X_i\) by
\(X_i^{(\ell)}\), where
\[
    X_i^{(\ell)}\sim F_i,
    \qquad i\in[n],\ \ell\in[k],
\]
and all \(nk\) variables are independent. The algorithm selects one
of these \(nk\) values, whereas the prophet is evaluated on a single copy of
the original instance.

For a fixed arrival model, \(M \geq 1\), and \(\epsilon>0\), the
\((1+\epsilon)\)-competition complexity is the smallest integer \(k\) such
that, for every instance \(F\) for which \(M(F)\) is at most $M$, there exists an
online algorithm using \(k\) copies and satisfying
\[
    \expect{ALG}
    \leq
    (1+\epsilon)\expect{OPT}.
\]
The arrival order of the \(nk\) variables depends on the model considered
below.
\paragraph{Block model.}
In the block model, the online algorithm observes \(k\) independent copies of
the original instance, one after another. For each block
\(\ell\in[k]\), the algorithm observes one independent copy
\(X_i^{(\ell)}\sim F_i\) of every variable \(X_i\), where all \(nk\) random
variables are independent. The variables within each block may arrive in an
arbitrary order, and this order may differ across blocks. Each copy is
referred to as a \emph{block}. We compare the expected cost of the online
algorithm on these \(k\) blocks with the expected minimum of a single copy of
the original instance.

\paragraph{Prophet secretary model.}
In the prophet secretary model, the algorithm observes \(k\) independent
copies of each variable, arriving in a uniformly random order, where all \(nk\) random variables are independent and
\(X_i^{(\ell)}\sim F_i\). A permutation
\(\pi\) is drawn uniformly at random from all permutations of these \(nk\)
variables, independently of their realizations, and the algorithm observes
the variables according to \(\pi\). Upon each arrival, the algorithm observes
both the identity of the variable and its realized value. The expectation of
the algorithm is taken over the realizations of the variables, the random
arrival order, and any internal randomness of the algorithm.

\paragraph{Adversarial-order model.}
In the adversarial-order model, an adversary chooses an arbitrary permutation
of the \(nk\) variables. The order may depend on the distributions
\(F_1,\ldots,F_n\), but it is fixed before the values are realized. The
algorithm must satisfy its guarantee for every such arrival order.

%% file: resource_augmentation/lower_bound_adversarial_order.tex
\section{Lower bound for adversarial 
order}
\label{lower bound adversarial order}
We now show that, in adversarial order, no finite amount of augmentation
guarantees a constant competitive ratio. More precisely, for every finite
number of copies \(k\), there are instances on which every online algorithm
has an arbitrarily large competitive ratio.

\begin{theorem}
\label{thm: adversarial lower bound}
For every \(k\ge 1\) and every \(M\ge2\), there exists an instance with
\[
    \mathbb{E}\big[\min\{X_1,X_2\}\big]=1
\]
such that, even after taking \(k\) copies of each variable, every online
algorithm with finite expected cost has expected cost at least \(M\) under an
adversarial arrival order.
\end{theorem}

\begin{proof}
See Appendix \ref{app: proof adversarial}.
\end{proof}

%% file: resource_augmentation/Algorithm_for_random_order.tex
\section{Algorithm for random order}
\label{Algortihm random order}
In this section, we show that \(O(\log \frac{M}{\epsilon})\) copies of each variable are enough to
obtain a $(1+\epsilon)$-competitive ratio in random order.
Denote $Y=\min_{i\in[n]} X_i $. Without loss of generality we assume $X_1$ is the variable with the minimum expectation, and hence $\mathbb{E}[X_1] =M\cdot\mathbb{E}[Y].$

For a given $k$, we consider the instance obtained by taking \(k\) independent copies of each
variable \(X_i\), and presenting all \(nk\) variables in a uniformly random order.

The algorithm is as follows.
Let \(\tau\) satisfy
\[
    \Pr\left[Y\le\tau\right]=\frac12.
\]
The algorithm accepts the first value \(x\le\tau\) that arrives before the
last copy of \(X_1\); if no such value appears, it accepts that last copy of
\(X_1\).
We now prove that this algorithm has expected cost at most $(1+\epsilon) \cdot \mathbb{E}[Y]$ for the appropriate $k$.

\begin{theorem}
Let $ k=16\lceil \log \frac{3M}{\epsilon}\rceil.$ The algorithm described above with $k$ independent copies of the original instance in random order satisfies
\[
    \mathbb{E}[\mathrm{ALG}] \le (1+\epsilon)\mathbb{E}\left[Y\right]
\]
Hence, the algorithm is \(1+\epsilon\)-competitive.
\end{theorem}

\begin{proof}
Denote the event that the algorithm takes the last copy of $X_1$ as $B$ and the event that the algorithm stops before the last copy of $X_1$ as $\overline{B}$.
The expectation of the algorithm is 
\[
\mathbb{E}[ALG] = \Pr(B)\cdot\mathbb{E}[ALG|B]+\Pr(\overline{B})\cdot\mathbb{E}[ALG|\overline{B}].
\]
The proof consists of two stages:
\begin{itemize}
    \item First, we show that
    \[
        \Pr(B)\mathbb{E}[\mathrm{ALG}\mid B]\le\epsilon\mathbb{E}[Y].
    \]
    \item Second, we show that
    \[
        \Pr(\overline B)
        \mathbb{E}[\mathrm{ALG}\mid\overline B]\le \mathbb{E}[Y].
    \]
\end{itemize}
\begin{claim}
\label{clm: 6.2}
    $\Pr(B)\cdot\mathbb{E}[ALG|B]\le \epsilon\mathbb{E}[Y]$.
\end{claim}
\begin{proof}
Let
\[
    L=\lceil \log \frac{3M}{\epsilon}\rceil,
    \qquad
    k=16L.
\]

We prove the claim by establishing the following two bounds:
\begin{enumerate}
    \item with probability at least
    \(1-\frac{\epsilon}{3M}\), the last copy of \(X_1\) appears in
    the final \(nL\) positions;
    \item with probability at least
    \(1-\frac{2\epsilon}{3M}\), some observation of value at most
    \(\tau\) appears among the first \(n(k-L)\) positions.
\end{enumerate}

\paragraph{The last copy of \(X_1\).}

We divide the random order into two parts. The prefix consists of the first
\(n(k-L)\) observations, and the suffix consists of the last \(nL\) observations.
 The bad event is that all \(k\) copies of \(X_1\) appear in the prefix. If \(n(k-L)<k\), this event is impossible and the desired bound is immediate. Otherwise, since the arrival order is uniformly random, the probability
that all \(k\) copies of \(X_1\) lie in the prefix is
\[
    \frac{\binom{n(k-L)}{k}}{\binom{nk}{k}} =
    \prod_{j=0}^{k-1}
    \frac{n(k-L)-j}{nk-j}.
\]
For every \(j\), we have
\[
    \frac{n(k-L)-j}{nk-j}
    \le
    \frac{n(k-L)}{nk}
    =
    1-\frac{L}{k}
    =
    1-\frac{1}{16}.
\]
Therefore,
\[
    \Pr[\text{all copies of }X_1\text{ appear in the prefix}]
    \le
    \left(1-\frac{1}{16}\right)^k
    \le
    e^{-k/16}
    =
    e^{-L}
    \le
    \frac{\epsilon}{3M}.
\]
Thus, with probability at least \(1-\frac{\epsilon}{3M}\), the last copy of \(X_1\)
appears in the suffix.

\paragraph{An eligible observation in the prefix.}
First we show that with high probability, there are many observations whose value
is at most \(\tau\) in the $nk$ variables. For every copy index \(t\in[k]\), define
\[
    Y_t=\min_{i\in[n]} X_i^{(t)},
\]
where \(X_i^{(t)}\) is the \(t\)-th independent copy of \(X_i\). Since
$Y_t$ has the same distribution as $Y$,
by definition of $\tau$ we get $\Pr[Y_t<\tau]= \frac{1}{2}.$
Let
\[
    N_{\tau}=\sum_{t=1}^{k}\mathbf{1}_{\{Y_t\le \tau\}}.
\]
Then    $\mathbb{E}[N_{\tau}]= \frac{k}{2}.$
Using the Chernoff bound,
\[
    \Pr\left[N_{\tau}\le \frac{k}{4}\right]
    \le
    \exp\left(-\frac{k}{16}\right)= \exp\left(-\frac{16L}{16}\right)= e^{-L}\le\frac{\epsilon}{3M}
\]
Hence, with probability at least \(1-\frac{\epsilon}{3M}\), there are at least \(k/4\)
copy indices \(t\) for which \(Y_t\le \tau\). In particular, there are at least
\(k/4\) observations whose value is at most \(\tau\).

Now condition on the event that there are at least \(k/4\) observations with
value at most \(\tau\). Since the order is uniformly random, these observations are
placed uniformly at random among the \(nk\) positions. The probability that all
of them lie in the suffix is at most
\[
    \left(\frac{nL}{nk}\right)^{k/4}
    =
    \left(\frac{1}{16}\right)^{k/4}\le \left(\frac{1}{16}\right)^{4\log(3M/\epsilon)}\le \frac{\epsilon}{3M}.
\]

Thus, with probability at least \(1-\frac{\epsilon}{3M}\) given that there are at least $\frac{k}{4}$ observations with value at most $\tau$, at least one of those observations appears in the prefix. Hence, by the union bound, we get that with probability at least $1-\frac{2\epsilon}{3M}$ a variable with value $\le \tau$ appears before the last $nL$ variables.
Therefore, again by the union bound, with probability at least \(1-\epsilon/M\), the  two things hold simultaneously: \begin{enumerate} 
\item the last copy of \(X_1\) appears in the suffix; 
\item some observation with value at most \(\tau\) appears in the prefix. 
\end{enumerate} On this event, the algorithm accepts a value at most \(\tau\) before it reaches the last copy of
\(X_1\). Hence, with probability at most $\frac{\epsilon}{M}$ the algorithm takes the last copy of $X_1$.

On the event $B$ the algorithm selects a copy of  $X_1$ independently of its value and $\mathbb{E}[X_1]=\mathbb{E}[Y]\cdot M$, we have  
\[
    \Pr(B)\cdot\mathbb{E}[ALG|B]=\Pr(B)\cdot\mathbb{E}[X_1]\le \frac{\epsilon}{M}\cdot M \cdot \mathbb{E}[Y] =\epsilon\mathbb{E}[Y].
\]
\end{proof}

We now bound the expected contribution from the event where the algorithm accepts a value at most $\tau$. 
\begin{claim}
\label{clm: 6.3}
    $ \Pr(\overline B)
        \mathbb{E}[\mathrm{ALG}\mid\overline B]\le\mathbb{E}[Y].$
\end{claim}

\begin{proof}
Consider the auxiliary algorithm that scans all \(nk\) observations in the
same random order as the real algorithm and accepts the first value at most
\(\tau\), if such a value exists. Let \(S\) be the event that the auxiliary
algorithm stops, and let \(Z\) be the value it accepts on \(S\).

We first show, by stochastic dominance, that
$\mathbb{E}[Z\mid S]\le \mathbb{E}[Y].$
Before determining the arrival order, label the \(k\) independent copies of
\(X_i\) by
\[
    X_i^{(1)},\ldots,X_i^{(k)},
\]
where \(X_i^{(\ell)}\sim F_i\). The superscript \(\ell\) identifies the copy and does not indicate
its position in the arrival order. After the values are realized, the \(nk\)
labeled observations are presented according to a uniformly random
permutation.
Let \(R\) denote the total number of observations, among all \(nk\)
observations, whose realized values are at most \(\tau\). The event \(S\) is
exactly the event \(\{R\ge 1\}\). Since $\Pr[Y>\tau]=\frac12,$
each block contains no value at most \(\tau\) with probability \(1/2\), and therefore $\Pr(S)=1-2^{-k}$. 

Fix \(a\le \tau\), and for every \(i\in[n]\), define
\[
    p_i(a)=\Pr[X_i\le a],
    \qquad
    p_i(\tau)=\Pr[X_i\le \tau].
\]
Fix one copy \(X_i^{(\ell)}\) of \(X_i\). Conditional on all realized values
and on \(X_i^{(\ell)}\le a\), the observation \(X_i^{(\ell)}\) is selected
precisely when it appears first among the \(R\) observations whose values
are at most \(\tau\). Since the arrival order is uniform, this occurs with
probability \(1/R\). Hence,
\begin{align*}
&\Pr\!\left[
    X_i^{(\ell)}\le a,\,
    X_i^{(\ell)}\text{ is selected}
    \,\middle|\, S
\right] \\
&\qquad=
\frac{p_i(a)}{1-2^{-k}}
\mathbb{E}\!\left[
    \frac1R
    \,\middle|\,
    X_i^{(\ell)}\le a
\right].
\end{align*}

The function \(x\mapsto 1/x\) is convex on \((0,\infty)\), so Jensen's
inequality gives
\[
\mathbb{E}\!\left[
    \frac1R
    \,\middle|\,
    X_i^{(\ell)}\le a
\right]
\ge
\frac{1}{
    \mathbb{E}[R\mid X_i^{(\ell)}\le a]
}.
\]
Moreover,
\[
\mathbb{E}[R\mid X_i^{(\ell)}\le a]
=
1+k\sum_{j=1}^n p_j(\tau)-p_i(\tau)
\le
1+k\sum_{j=1}^n p_j(\tau).
\]
Consequently,
\[
\Pr\!\left[
    X_i^{(\ell)}\le a,\,
    X_i^{(\ell)}\text{ is selected}
    \,\middle|\, S
\right]
\ge
\frac{p_i(a)}
{(1-2^{-k})
 \left(1+k\sum_{j=1}^n p_j(\tau)\right)}.
\]

The events in which the auxiliary algorithm selects different observations
are disjoint. Therefore, summing over all \(k\) copies and all \(i\in[n]\)
yields for any \(a\le \tau\) the following
\[
\Pr[Z\le a\mid S]
\ge
\frac{
    k\sum_{i=1}^n p_i(a)
}{
    (1-2^{-k})
    \left(1+k\sum_{i=1}^n p_i(\tau)\right)
}.
\]

It remains to bound \(\sum_i p_i(\tau)\). Since
\[
    \prod_{i=1}^n(1-p_i(\tau))
    =
    \Pr[Y>\tau]
    =
    \frac12,
\]
the Arithmetic-Geometric mean inequality implies
\[
1-\frac1n\sum_{i=1}^n p_i(\tau)
=
\frac1n\sum_{i=1}^n(1-p_i(\tau))
\ge
\left(\prod_{i=1}^n(1-p_i(\tau))\right)^{1/n}
=
2^{-1/n} \ge
    1-\frac{\ln 2}{n}
\]
where the last inequality follows from \(e^{-x}\ge 1-x\).
Hence
\[
    \sum_{i=1}^n p_i(\tau)\le \ln 2.
\]
It follows that
\[
\Pr[Z\le a\mid S]
\ge
\frac{k}
{(1-2^{-k})(1+k\ln 2)}
\sum_{i=1}^n p_i(a).
\]
It is easy to verify that for every \(k\ge1\),
\[
\frac{k}
{(1-2^{-k})(1+k\ln 2)}
\ge 1.
\]
Thus, for any \(a\le \tau\) we have
\[
    \Pr[Z\le a\mid S]
    \ge
    \sum_{i=1}^n p_i(a)
    \ge
    \Pr[Y\le a],
\]
where the last inequality follows from the union bound.

For \(a>\tau\), the inequality above is immediate, since \(Z\le\tau\) on \(S\).
Therefore,
\[
    \Pr[Z\le a\mid S]= 1\ge \Pr[Y\le a]
\]
for every \(a\ge0\). Thus, $Y$ stochastically dominates  \(Z\mid S\).
Since both variables are nonnegative,
\[
\mathbb{E}[Z\mid S]
=
\int_0^\infty \Pr[Z>a\mid S]\,da
\le
\int_0^\infty \Pr[Y>a]\,da
=
\mathbb{E}[Y]
.
\]

We now return to the real algorithm. Let \(\overline B\) be the event that
the real algorithm accepts a value at most \(\tau\) before the last copy of
\(X_1\). On \(\overline B\), the real and auxiliary algorithms select the
same observation, and \(\overline B\subseteq S\). Therefore,
\[
\Pr(\overline B)\mathbb{E}[\mathrm{ALG}\mid\overline B]
=\mathbb{E}\!\left[\mathrm{ALG}\cdot\mathbf{1}_{\overline B}\right]
=\mathbb{E}\!\left[Z\cdot\mathbf{1}_{\overline B}\right]
\le \mathbb{E}\!\left[Z\cdot\mathbf{1}_{S}\right]
=\Pr(S)\mathbb{E}[Z\mid S]\le \mathbb{E}[Y].
\]
This proves the claim.
\end{proof}

Combining Claims~\ref{clm: 6.2} and~\ref{clm: 6.3}, and partitioning according
to whether the algorithm stops upon observing a value at most \(\tau\) or
selects the last copy of \(X_1\), we obtain
\[
\mathbb{E}[ALG] \le \Pr(B)\cdot\mathbb{E}[ALG|B]+\Pr(\overline{B})\cdot\mathbb{E}[ALG|\overline{B}]\le (1+\epsilon)\mathbb{E}[Y],
\]
which proves the theorem. \end{proof}

%% file: resource_augmentation/lower_bound_for_random_order.tex
\section{Lower bound for random order}
\label{lower bound random order}
We now show that for $M\ge2$ at least \(\Omega(\log \frac{M}{\epsilon})\) copies are necessary to obtain a \(1+\epsilon\) competitive ratio in random order. 

\begin{theorem}
Let \(\epsilon\in(0,1)\), \(M\ge2\), and 
$ k\le \frac12\log\left(\frac{M}{4\epsilon}\right)$.
Then there exists an instance with
\[
    \mathbb{E}[\min\{X_1,X_2\}]=1
    \qquad\text{and}\qquad
    \min_{i\in\{1,2\}}\mathbb{E}[X_i]=M
\]
such that every online algorithm on $k$ copies of the instance that arrive in a random order 
has expected cost
strictly larger than \(1+\epsilon\).
Consequently, achieving a \(1+\epsilon\)-competitive ratio requires
$  \Omega\!\left(\log\frac{M}{\epsilon}\right) $ copies.

\end{theorem}
\begin{proof}
Fix \(\epsilon\in(0,1)\), \(M\ge2\), and \(k\ge1\). Consider
\[
    X_1\equiv M,
\]
and
\[
    X_2=
    \begin{cases}
    1-\epsilon
    & \text{with probability } \dfrac{M-1}{M-1+\epsilon},\\[6pt]
    B
    & \text{with probability } \dfrac{\epsilon}{M-1+\epsilon},
    \end{cases}
\]
where
\[
    B
    =
    M\left(\frac{M-1+\epsilon}{\epsilon}\right)^k.
\]
For a single copy of each variable, we have
\[
\begin{aligned}
    \mathbb{E}[\min\{X_1,X_2\}]
    &=
    \frac{M-1}{M-1+\epsilon}(1-\epsilon)
    +
    \frac{\epsilon}{M-1+\epsilon}M \\
    &=
    \frac{(M-1)(1-\epsilon)+\epsilon M}{M-1+\epsilon} \\
    &=
    \frac{M-1+\epsilon}{M-1+\epsilon} =1.
\end{aligned}
\]
Also,
\[
\mathbb{E}[X_2] \ge B\cdot\left(\dfrac{\epsilon}{M-1+\epsilon}\right)= M\left(\frac{M-1+\epsilon}{\epsilon}\right)^k\left(\dfrac{\epsilon}{M-1+\epsilon}\right)\ge M.
\]
Thus, \(\min_i\mathbb{E}[X_i]=M\).

Now take \(k\) independent copies of each variable, presented in a uniformly
random order. Let \(E\) be the event that all copies of \(X_1\) arrive before
all copies of \(X_2\). Then
\[
    \Pr[E]=\frac{1}{\binom{2k}{k}}>4^{-k}.
\]

Conditioned on \(E\), if the algorithm accepts one of the first \(k\) values,
its cost is \(M\). If it rejects all of them, then even if it could observe all
remaining values and select their minimum, its expected cost would be at least
\[
    \mathbb{E}\!\left[\min_{\ell\in[k]}X_2^{(\ell)}\right]
    \ge
    \left(\frac{\epsilon}{M-1+\epsilon}\right)^k B
    =M.
\]
Thus, \(\mathbb{E}[\mathrm{ALG}\mid E]\ge M\). Moreover, every value in the
support of the instance is at least \(1-\epsilon\), so
\(\mathbb{E}[\mathrm{ALG}\mid\overline E]\ge1-\epsilon\). Therefore,
\[
\begin{aligned}
    \mathbb{E}[\mathrm{ALG}]
    &\ge M\Pr[E]+(1-\epsilon)\Pr[\overline E] \\
    &=(M-1+\epsilon)\Pr[E] + 1-\epsilon \\
    &>(M-1+\epsilon)4^{-k} + 1-\epsilon.
\end{aligned}
\]

By the assumption that $ k\le \frac12\log\left(\frac{M}{4\epsilon}\right)$ we have
    $4^{-k}\ge\frac{4\epsilon}{M}$.
Therefore,
\[
\begin{aligned}
    \mathbb{E}[\mathrm{ALG}]
    &>
    1-\epsilon+(M-1+\epsilon)4^{-k} \\
    &\ge
    1-\epsilon+(M-1+\epsilon)
    \frac{4\epsilon}{M}\\
    &> 1-\epsilon+2\epsilon=
    1+\epsilon.
\end{aligned}
\]
The last inequality follows because $M\ge 2$.
Therefore, achieving a \(1+\epsilon\)-competitive ratio requires
\( \Omega\!\left(\log\frac{M}{\epsilon}\right) \)
copies in this construction,
which completes the proof.
\end{proof}

%% file: resource_augmentation/algorithm_for_loglogM_interleaved_variables.tex
\section{Block model: constant competitive algorithm}
\label{constant block algorithm}
Before turning to the \((1+\epsilon)\)-competitive algorithm, we give a
simple constant-competitive algorithm with weaker information requirements.
Unlike the algorithm of Section \ref{block algorithm}, which requires access to quantiles of the
distribution of the minimum, this algorithm uses only the expectations of the individual variables and their minimum.
The algorithm  achieves a constant competitive ratio using
\(O(\log\log M)\) blocks.  Without loss of generality, we assume $X_1$ is the variable with the minimum expectation, and hence $\mathbb{E}[X_1] =M\cdot\mathbb{E}[OPT].$

\begin{algorithm}[H]
\caption{\(\log\log M\)-block threshold rule}
\label{alg:constant-block}
\begin{enumerate}
    \item Set
    $g=\lceil\log\log M\rceil\qquad\text{and}\qquad T_j=2\cdot 2^{2^j} \mathbb{E}[OPT].
    $
    \item For each block \(j=1,\ldots,g-1\), accept the first value
    \(x_i\le T_j\), if one appears.
    \item If no value was accepted, accept the copy of \(X_1\) in block \(g\).
\end{enumerate}
\end{algorithm}

\begin{theorem}
For every instance of independent nonnegative random variables and large enough $M$, the above algorithm is constant-competitive in the block model using
   $ O(\log\log M) $
blocks.
\end{theorem}

\begin{proof}
Let \(R_i\) be the event that block \(i\) is reached. Denote $t_j= 2\cdot2^{2^j}$ and $\mu=\mathbb{E}[OPT]$. By independence and
Markov's inequality,
\[
    \Pr[R_i]
    \le
    \prod_{j=1}^{i-1}\frac{\mu}{T_j}=\prod_{j=1}^{i-1}\frac{1}{t_j}.
\]
The value selected in block \(i<g\) is at most \(T_i\). In the final block,
the algorithm selects an independent copy of \(X_1\), whose expectation is
\(M\cdot\mu\). Since \(g=\lceil \log\log M\rceil\), we have \(M\le 2^{2^g}<t_g\). Hence $\mathbb{E}[X_1]=M\mu\le \mu t_g=T_g,$
so we can bound the contribution of the final block in the same way as the preceding blocks. Therefore,
\[
\begin{aligned}
    \mathbb E[\mathrm{ALG}]
    &\le
   \sum_{i=1}^{g-1}
        T_i\prod_{j=1}^{i-1}\frac1{t_j}
    +\mathbb{E}[X_1]\prod_{j=1}^{g-1}\frac1{t_j}\\
    &\le
    \mu\cdot\sum_{i=1}^{g}
        t_i\prod_{j=1}^{i-1}\frac{1}{t_j}
    =
    \mu\sum_{i=1}^{g}2^{4-i}
    \le
    \mu\sum_{i=1}^{\infty}2^{4-i}
    =16\mu.
\end{aligned}
\]
Therefore, the algorithm is $ 16$-competitive. Finally,
\[
    g=\lceil\log\log M\rceil=O(\log\log M),
\]
which proves the claimed block complexity.
\end{proof}

%% file: resource_augmentation/Algortihm_i.i.d_loglog.tex
\section{Block model: a $(1+\epsilon)$-competitive algorithm}
\label{block algorithm}
In this section, we prove that a \((1+\epsilon)\)-competitive ratio can be
achieved in the block model using
   $ O\left(\log\log\frac{M}{\epsilon}\right) $
blocks. We first derive the conditional-expectation guarantee that we need
from the block-threshold maximization algorithm of
\cite{BrustleCDEFV26}. We then use this algorithm as a subroutine in a
reduction from minimization to maximization.

\begin{theorem}
\label{thm:block-upper-bound}
Let \(X_1,\ldots,X_n\) be independent nonnegative random variables.
For every \(\epsilon\in(0,1)\), there exists a
\((1+\epsilon)\)-competitive algorithm in the block model using
   $ O\left(\log\log\frac{M}{\epsilon}\right) $
blocks.
\end{theorem}
We first recall the block-threshold maximization algorithm of \cite{BrustleCDEFV26}. Given independent nonnegative random variables \(V_1,\ldots,V_n\), let \(V=\max_{i\in[n]}V_i\). In block \(\ell\), the algorithm uses a threshold \(\tau_\ell\) satisfying 
\( \Pr[V\ge\tau_\ell] = 1-\left(\frac54\right)^{-(5/4)^\ell}, \)
and accepts the first value exceeding the threshold of its block. For every \(\eta\in(0,1)\), using 
\( \max\left\{ 2, \left\lceil \log_{5/4}\log_{5/4}\frac1\eta \right\rceil \right\} \)
blocks, this algorithm achieves a \((1-\eta)\)-competitive ratio. We will use a slightly stronger consequence of their analysis below.\footnote{ More precisely, \cite{BrustleCDEFV26} defines the interval factors \(\Phi_1,\ldots,\Phi_{r+1}\) before Lemma~1, where \(r\) denotes the number of blocks. Lemma~1 bounds the tail probability of the algorithm separately on the intervals induced by its thresholds. The proof of Lemma~2 then shows that, for the quantiles used above, the corresponding factor is at least \(1\) on every interval above the final threshold, while on the last interval it is at least \(1-(5/4)^{-(5/4)^r}\).} For defining our minimization algorithm we use a dual algorithm.

\paragraph{Algorithm.} Let $ Y=\min_{i\in[n]}X_i,$ and, without loss of generality, assume that \(X_1\) has the smallest expectation, so that $ \mathbb E[X_1]=M\cdot\mathbb E[Y].$ Set 
\[
\delta=\frac{\epsilon}{M}, \qquad k=\max\left\{ 2, \left\lceil \log_{5/4}\log_{5/4}\frac1\delta \right\rceil \right\}. 
\]
For \(i\in[k+1]\), let \(\rho_i\) satisfy $ \Pr[Y\le\rho_i] = 1-\left(\frac54\right)^{-(5/4)^i}$. 

\begin{algorithm}[H]  \label{alg:block-model}
\caption{Block-model algorithm}
\begin{enumerate} 
\item Accept the first arriving variable in block $i$ whose value its value is at most 
\(\rho_i\).
\item If a variable has not been selected yet then select an independent copy of \(X_1\) from one additional block. 
\end{enumerate} 
\end{algorithm} 

Before analyzing Algorithm \ref{alg:block-model} we set $ B=\rho_{k+1}. $ For every \(i\in[n]\), define 
\[
X_i^B=\min\{X_i,B\}, \qquad W_i=B-X_i^B, \qquad D=\max_{i\in[n]}W_i. \] 

We observe that \( D = B-\min\{Y,B\}.  \) 
For every \(\ell\in[k]\), set $ \tau_\ell=B-\rho_\ell. $ Since \(\rho_\ell<B\), we have \(\tau_\ell>0\). Moreover, 
\[ \Pr[D\ge\tau_\ell] = \Pr\!\left[B-\min\{Y,B\}\ge B-\rho_\ell\right] = \Pr[Y\le\rho_\ell] = 1-\left(\frac54\right)^{-(5/4)^\ell}. 
\]
Thus, these are precisely the thresholds used by the block-threshold maximization algorithm. Our analysis is actually done by a reduction to the maximization algorithm.

Note that the first phase never selects a value larger than \(B\), since such a value exceeds every threshold used in this phase. Hence, whenever \(X_i\) is selected,
\[
W_i=B-X_i^B=B-X_i.
\]
Therefore, under the transformation \(W_i=B-X_i^B\), the first phase of our algorithm corresponds exactly to running the maximization dual algorithm on the variables \(W_i\).

\paragraph{Analysis of the first phase.} Let 
\[
\gamma_k=\left(\frac54\right)^{-(5/4)^k}. 
\] 
Let $V$ be the variable selected on the first phase of the minimization algorithm and let $W$ be the selected variable  in the maximization problem. Clearly $W=B-V$.
The interval-by-interval analysis in the proofs of Lemmas~1 and~2 of \cite{BrustleCDEFV26} implies that, for every \(x>0\), 
\begin{equation}
\Pr[W\ge x] \ge \begin{cases} \Pr[D\ge x], & x\ge\tau_k,\\[2mm] (1-\gamma_k)\Pr[D\ge x], & 0<x<\tau_k. \end{cases} \label{eq:block-tail-guarantee} 
\end{equation} 
Thus, the only loss in the tail guarantee occurs below the final threshold. Since every value selected by the maximization algorithm is at least \(\tau_k\), conditioning on the algorithm stopping eliminates this loss. The following lemma formalizes this observation. 
\begin{lemma}
\label{lem:block-maximization} 
Let \(S\) be the event that the first phase stops. 
Then \[ \Pr[S]\ge1-\delta \qquad\text{and}\qquad \mathbb E[V\mid S]\le\mathbb E[Y]. \] 
\end{lemma}
\begin{proof}
We actually analyze the dual algorithm. Let  $p_\ell=\Pr[D\ge\tau_\ell] = 1-\left(\frac54\right)^{-(5/4)^\ell}. $ A fresh block causes the algorithm to stop in block \(\ell\) with probability \(p_\ell\), conditional on reaching that block. Since the blocks are independent, 
\[
 \Pr[\overline S] = \prod_{\ell=1}^k(1-p_\ell) \le 1-p_k = \left(\frac54\right)^{-(5/4)^k} \le\delta, 
 \]
 where the last inequality follows from the definition of \(k\). Thus, \(\Pr[S]\ge1-\delta\). We now prove the conditional-expectation guarantee. Since the thresholds are decreasing, every value selected by the maximization phase is at least \(\tau_k\). Therefore, for every \(0\le x<\tau_k\), 
 \[ 
 \Pr[W\ge x\mid S] = 1 \ge \Pr[D\ge x].
 \] 
 Now consider \(x\ge\tau_k\). Since \(\{W\ge x\}\subseteq S\), by \eqref{eq:block-tail-guarantee},
 \[
 \Pr[W\ge x\mid S] = \frac{\Pr[W\ge x]}{\Pr[S]}\ge \Pr[W\ge x]\ge \Pr[D\ge x]. 
 \] 
 Hence, conditioned on \(S\), the selected value \(W\) stochastically dominates \(D\). Integrating the tail probabilities gives \[ 
 \mathbb E[W\mid S] = \int_0^\infty \Pr[W\ge x\mid S]\,dx\ge \int_0^\infty \Pr[D\ge x]\,dx= \mathbb E[D].
 \] 

 Finally, using the above inequality we get
 \begin{align*} 
  \mathbb E[V\mid S] & =  \mathbb E[B-W\mid S] = B-\mathbb E[W\mid S] \\
  &\le B-\mathbb E[D]= B- \mathbb E[\max_i\{B-X_i^B\}]  = \mathbb E[\min_iX_i^B] \le \mathbb E[\min_iX_i] \le \mathbb E[Y].      
 \end{align*}
 
 \end{proof} 
 \begin{proof}[Proof of Theorem~\ref{thm:block-upper-bound}] 
  
  On \(\overline S\), the algorithm selects an independent copy of \(X_1\), whose expected value is 
  \[ 
  \mathbb E[X_1]=M\cdot\mathbb E[Y].
  \]
  By Lemma \ref{lem:block-maximization} we have \(\Pr[\overline S]\le\delta\) and \(\mathbb E[\mathrm{ALG}\mid S] \le \mathbb E[Y] \). Hence 
  \[  \mathbb E[\mathrm{ALG}] = \Pr[S]\mathbb E[\mathrm{ALG}\mid S] + \Pr[\overline S]\mathbb E[X_1]\le \mathbb E[Y] + \delta M\mathbb E[Y]= (1+\epsilon)\mathbb E[Y],
  \] 
  where the last equality follows from \(\delta=\epsilon/M\). Finally, the first phase uses 
  \(
  O\left(\log\log\frac1\delta\right) = O\left(\log\log\frac{M}{\epsilon}\right)
  \) 
  blocks. Together with the additional fallback block, the total number of blocks remains 
  \(
  O\left(\log\log\frac{M}{\epsilon}\right).
  \) 
  This proves the theorem. \end{proof}

%% file: resource_augmentation/Block_model_Lower_Bound.tex
\section{Lower bound for the block model}
\label{sec:block-lower-bound}

\begin{theorem}
\label{thm:block-lower-bound}
For all sufficiently large \(M\) and sufficiently small
\(\epsilon>0\), the competition complexity of the i.i.d.\ block model is
\[
    \Omega\left(\log\log\frac{M}{\epsilon}\right).
\]
\end{theorem}

To prove the theorem we establish the dependence on \(M\) and on \(\epsilon\) separately.
First, we construct an i.i.d.\ instance for which
\(\Omega(\log\log M)\) blocks are necessary even to obtain a constant
competitive ratio. We then apply an affine transformation to the same hard
distribution and obtain an instance with parameter $M$ at most \(2\) for which
\(\Omega(\log\log(1/\epsilon))\) blocks are necessary to obtain a
\((1+\epsilon)\)-competitive ratio.

Since competition complexity is a worst-case quantity, the two lower bounds
combine to give
\[
    \Omega\left(
        \max\left\{
            \log\log M,
            \log\log\frac1\epsilon
        \right\}
    \right)
    =
    \Omega\left(\log\log\frac{M}{\epsilon}\right).
\]

\begin{lemma}[Dependence on \(M\)]
\label{lem:block-lower-M}
For every \(M\) which is not too small, there exists an i.i.d.\
minimization instance
such that every \(2\)-competitive online algorithm requires
   $ \Omega(\log\log M) $
blocks.
\end{lemma}

\begin{proof}
Fix \(n\ge 4\) and let
$\alpha=\frac{2}{n}.$
Fix some \(R>1\), and set $r=R^{1/(1-\alpha)}.$
Let \(X\) be distributed according to
\[
F(x)=
\begin{cases}
0, & x<1,\\[1mm]
1-\dfrac{1}{x^\alpha}, & 1\le x<r,\\[2mm]
1, & x\ge r.
\end{cases}
\]
Thus, \(X\) has an atom of mass \(r^{-\alpha}\) at \(r\).

We first calculate the expectation of one variable and the expected offline
optimum. These calculations will also relate the auxiliary parameter \(R\)
to the parameter \(M\) of the instance. Using the tail-integral formula,
\begin{align*}
    \mathbb E[X]
    &=
    1+\int_1^r \Pr[X>x]\,dx 
    =
    1+\int_1^r x^{-\alpha}\,dx \\
    &=
    \frac{r^{1-\alpha}-\alpha}{1-\alpha}
    =
    \frac{R-\alpha}{1-\alpha}.
\end{align*}
Since \(R\ge1\) and \(\alpha\le1/2\), it follows that
\[
    R\le \mathbb E[X]\le 2R.
\]

Let \(X_1,\ldots,X_n\) be independent copies of \(X\), and denote
$ Y=\min_{i\in[n]}X_i.$ For \(1\le x<r\),
\[
    \Pr[Y>x]
    =
    \Pr[X>x]^n
    =
    x^{-n\alpha}
    =
    x^{-2}.
\]
Hence,
\begin{align*}
    \mathbb E[Y]
    =
    1+\int_1^r \Pr[Y>x]\,dx 
    =
    1+\int_1^r x^{-2}\,dx 
    =
    1+\left(1-r^{-1}\right)
    \le 2.
\end{align*}
Moreover, \(Y\ge1\), and hence \(\mathbb E[Y]\ge1\). Since the instance
is i.i.d., its parameter is
$M=\frac{\mathbb E[X]}{\mathbb E[Y]}.$ Combining the preceding bounds implies that
\[
    \frac{R}{2}\le M\le 2R.
\]

\paragraph{The optimal online cost.}
For \(t\ge0\), let \(V_t\) denote the optimal expected cost when \(t\)
i.i.d.\ observations remain, where \(V_0=\infty\). By backward
induction,
\[
    V_{t+1}
    =
    \mathbb E\left[\min\{X,V_t\}\right].
\]
Indeed, upon observing \(X\), the optimal algorithm accepts it if
\(X\le V_t\), and otherwise continues with the remaining \(t\)
observations.

We claim that, for every \(t\ge0\),
\[
    V_t\ge R^{(1-\alpha)^t}.
\]
The claim is immediate for \(t=0\), since $V_0 =\infty\ge R$.  Suppose it holds for \(t\), and set
$L=R^{(1-\alpha)^t}.$ Since \(V_t\ge L\),
\begin{align*}
    V_{t+1}
    &=
    \mathbb E[\min\{X,V_t\}] 
    \ge
    \mathbb E[\min\{X,L\}]\\ 
    &=
    1+\int_1^L x^{-\alpha}\,dx 
    =
    \frac{L^{1-\alpha}-\alpha}{1-\alpha}
    \ge
    L^{1-\alpha} 
    =
    R^{(1-\alpha)^{t+1}}.
\end{align*}
This proves the claim.

Now let
\[
    k=\left\lfloor\frac18\log\log R\right\rfloor
\]
and consider an algorithm using \(k\) blocks. Since each block contains
\(n\) i.i.d.\ observations, the algorithm observes
\[
    N=kn\le\frac{n\log\log R}{8}
\]
i.i.d.\ variables in total. Hence,
\[
    \mathbb E[\mathrm{ALG}]
    \ge
    V_N
    \ge
    R^{(1-\alpha)^N}.
\]
Using \(\alpha=2/n\) and
\[
    \left(1-\frac{2}{n}\right)^{n/2}\ge\frac14
    \qquad\text{for }n\ge 4,
\]
we obtain
\begin{align*}
    (1-\alpha)^N
    &\ge
    \left(1-\frac{2}{n}\right)^{
        \frac{n\log\log R}{8}
    } 
    =
    \left(
        \left(1-\frac{2}{n}\right)^{n/2}
    \right)^{\frac{\log\log R}{4}} \\
    &\ge
    \left(\frac14\right)^{\frac{\log\log R}{4}}
    =
    \frac{1}{\sqrt{\log R}}.
\end{align*}
Therefore,
\[
    \mathbb E[\mathrm{ALG}]
    \ge
    R^{1/\sqrt{\log R}}
    =
    2^{\sqrt{\log R}},
\]
where logarithms are base \(2\). On the other hand,
$ \mathbb E[\mathrm{OPT}]=\mathbb E[Y]\le 2.$
Thus, the competitive ratio of even the optimal online algorithm using
\(k\) blocks is at least
\[
    \frac{2^{\sqrt{\log R}}}{2},
\]
which is larger than \(2\) for all sufficiently large \(R\). Hence every
\(2\)-competitive algorithm requires
$\Omega(\log\log R)$ blocks. Finally, since \(M=\Theta(R)\), this is also
\(\Omega(\log\log M)\).
\end{proof}
\begin{lemma}[Dependence on \(\epsilon\)]
\label{lem:block-lower-epsilon}
For every sufficiently small \(\epsilon>0\), there exists an i.i.d.\
minimization instance with parameter $M$ at most \(2\) such that every
\((1+\epsilon)\)-competitive online algorithm requires
\(
    \Omega\left(\log\log\frac1\epsilon\right)
\)
blocks.
\end{lemma}

\begin{proof}
Fix \(n\ge 4\) and let $\alpha=\frac{2}{n}.$ Set
\[
    r_\epsilon=1+\frac1\epsilon
    \qquad\text{and}\qquad
    R_\epsilon=r_\epsilon^{1-\alpha}.
\]
Consider \(n\) i.i.d.\ random variables with CDF
\[
G(x)=1-\frac{1}{x^\alpha},
\qquad
x\in[1,r_\epsilon),
\]
and an atom of mass \(r_\epsilon^{-\alpha}\) at \(r_\epsilon\).

This is precisely the hard distribution from
Lemma~\ref{lem:block-lower-M}, instantiated with \(R=R_\epsilon\), since
\[
    r_\epsilon=R_\epsilon^{1/(1-\alpha)}.
\]
Therefore, every \(2\)-competitive algorithm for this instance requires
$\Omega(\log\log R_\epsilon)$ blocks. Moreover,
\[
\begin{aligned}
    \log\log R_\epsilon
    =
    \log\left((1-\alpha)\log\left(1+\frac1\epsilon\right)\right) 
    =
    \Theta\left(\log\log\frac1\epsilon\right).
\end{aligned}
\]

Let \(X\) denote a random variable with CDF \(G\), and define
 $Z=1+\epsilon(X-1).$
This increasing affine transformation maps
\([1,1+1/\epsilon]\) onto \([1,2]\). In particular, \(Z\in[1,2]\).
Thus,
\[
    \mathbb E[Z]\le2
    \qquad\text{and}\qquad
    \mathbb E\left[\min_{i\in[n]}Z_i\right]\ge1,
\]
so the parameter $M$ of the transformed instance is at most \(2\).

Suppose that there exists a \((1+\epsilon)\)-competitive algorithm
\(A_Z\) for the transformed instance. We construct an algorithm \(A_X\)
for the original instance as follows. Whenever \(A_X\) observes \(X_j\),
it computes $Z_j=1+\epsilon(X_j-1)$ and feeds \(Z_j\) to \(A_Z\). The algorithm \(A_X\) stops exactly when
\(A_Z\) stops and selects the corresponding observation. Thus, the two
selected values satisfy
\[
    A_X=1+\frac{A_Z-1}{\epsilon}.
\]

Let
\[
    \mu_X=\mathbb E\left[\min_{i\in[n]}X_i\right],
    \qquad
    \mu_Z=\mathbb E\left[\min_{i\in[n]}Z_i\right].
\]
Since the transformation is increasing, $\mu_Z=1+\epsilon(\mu_X-1).$
Using the \((1+\epsilon)\)-competitive guarantee of \(A_Z\), we obtain
\begin{align*}
    \mathbb E[A_X]
    &=
    1+\frac{\mathbb E[A_Z]-1}{\epsilon} 
    \le
    1+\frac{(1+\epsilon)\mu_Z-1}{\epsilon} \\
    &=
    1+\frac{
        (1+\epsilon)(1+\epsilon(\mu_X-1))-1
    }{\epsilon} 
    =
    (1+\epsilon)\mu_X+1-\epsilon.
\end{align*}
Since \(\mu_X\ge1\) and \(\epsilon\le1\),
\[
    (1+\epsilon)\mu_X+1-\epsilon
    \le 2\mu_X.
\]
Hence \(A_X\) is \(2\)-competitive for the original hard instance.

The reduction preserves the block structure, so \(A_X\) and \(A_Z\)
use the same number of blocks. Lemma~\ref{lem:block-lower-M} therefore
implies that \(A_Z\) requires
\(
    \Omega(\log\log R_\epsilon)
    =
    \Omega\left(\log\log\frac1\epsilon\right)
\)
blocks.
\end{proof}

\begin{proof}[Proof of Theorem~\ref{thm:block-lower-bound}]
First, Lemma~\ref{lem:block-lower-M} yields a lower bound of
\(
    \Omega(\log\log M)
\)
on the number of blocks, even for a constant competitive ratio. In particular, this lower
bound applies for $\epsilon=\frac12$. 
In addition, 
Lemma~\ref{lem:block-lower-epsilon} yields a lower bound of
\(
    \Omega\left(\log\log\frac1\epsilon\right)
\)
on the number of blocks, already for instances whose parameter $M$ is at most \(2\).
Combining the two lemmas, the worst-case competition complexity is at least
\[
    \Omega\left(
        \max\left\{
            \log\log M,
            \log\log\frac{1}{\epsilon}
        \right\}
    \right)
    = 
    \Omega\left(
            \log\log M +
            \log\log\frac{1}{\epsilon}
    \right)
    =
     \Omega\left(\log\log\frac{M}{\epsilon}\right)
\]
\ignore{
Moreover,
\[
\begin{aligned}
    \log\log\frac{M}{\epsilon}
    =
    \log\left(
        \log M+\log\frac{1}{\epsilon}
    \right)
    =
    \Theta\left(
        \max\left\{
            \log\log M,
            \log\log\frac{1}{\epsilon}
        \right\}
    \right),
\end{aligned}
\]
where the last equality holds because
\[
    \max\left\{\log M,\log\frac{1}{\epsilon}\right\}
    \le
    \log M+\log\frac{1}{\epsilon}
    \le
    2\max\left\{\log M,\log\frac{1}{\epsilon}\right\}.
\]
Therefore, the competition complexity is
\[
    \Omega\left(\log\log\frac{M}{\epsilon}\right),
\]
}
which proves the theorem.
\end{proof}

%% file: sections/Conclusion.tex
\section{Conclusion}

We initiated the study of competition complexity for the single-choice min-prophet inequality. We showed that the $(1+\epsilon)$-competition complexity is $\Theta\!\left(\log\log\frac{M}{\epsilon}\right)$ in the block model and $\Theta\!\left(\log\frac{M}{\epsilon}\right)$ in the prophet secretary model, while under fully adversarial arrivals no finite number of copies guarantees a constant competitive ratio. These results highlight the min-prophet problem's strong dependence on the arrival model and show that resource augmentation can overcome the impossibility of constant-factor guarantees only when sufficient structure is present.

A natural open problem is to identify broad and natural families of distributions for which a finite number of copies suffices to obtain a $(1+\epsilon)$-competitive algorithm even under adversarial arrivals. 

%% file: Appendix/Proof_Adverserial_lower_bound.tex
\section{Proof of Theorem \ref{thm: adversarial lower bound}}
\label{app: proof adversarial}
\begin{proof}
Fix \(M\ge2\) and $k$. Consider two random variables:
\[
    X_1 \equiv M,
\]
and
\[
    X_2 =
    \begin{cases}
    0 & \text{with probability } 1-\frac{1}{M},\\[4pt]
    M^{k+1} & \text{with probability } \frac{1}{M}.
    \end{cases}
\]
For a single copy of each variable, we have
\[
    \mathbb{E}\big[\min\{X_1,X_2\}\big]
    =
    \left(1-\frac{1}{M}\right)\cdot 0
    +
    \frac{1}{M}\cdot M
    =
    1.
\]

Now take \(k\) independent copies of \(X_1\) and \(k\) independent copies of
\(X_2\). Consider the adversarial order in which all copies of \(X_1\) arrive
first, followed by all copies of \(X_2\):
\[
    \underbrace{X_1,\dots,X_1}_{k\text{ copies}},
    \underbrace{X_2,\dots,X_2}_{k\text{ copies}}.
\]

Consider any online algorithm, and let \(q\) be the probability that it accepts
one of the first \(k\) arrivals. On this event, its cost is \(M\). If it rejects
all \(k\) copies of \(X_1\), then it can select only among the \(k\) copies of
\(X_2\). Even if it were given the additional power to observe all of them and
select their minimum, its expected cost would be
\[
    \mathbb{E}\!\left[\min_{\ell\in[k]}X_2^{(\ell)}\right]
    =
    \left(\frac{1}{M}\right)^k M^{k+1}
    =
    M.
\]
Therefore,
\[
    \mathbb{E}[\mathrm{ALG}]
    \ge qM+(1-q)M
    =M.
\]

Since the prophet's expected cost is \(1\), the competitive ratio is at least
\(M\). Because \(M\) can be chosen arbitrarily large, no constant competitive
ratio is possible in adversarial order, regardless of the finite number of
copies \(k\).
\end{proof}